\documentclass[onecolumn,footinbib,floatfix,aps,10pt,notitlepage]{revtex4-1}

\usepackage{graphicx}
\usepackage{epsfig}
\usepackage{epstopdf}
\usepackage{amsmath}
\usepackage{amsfonts}
\usepackage{txfonts}
\usepackage{amssymb}
\usepackage{amstext}
\usepackage[sort&compress]{natbib}
\usepackage{rotating}
\usepackage{dcolumn}
\usepackage{graphics}
\usepackage{array,multirow}
\usepackage[english]{babel}
\usepackage{attachfile}
\usepackage{pgfplots}
\usepackage[export]{adjustbox}
\usepackage{braket}

\pgfplotsset{compat=1.13}

\def\be{\begin{equation}}
\def\ee{\end{equation}}
\def\bea{\begin{eqnarray}}
\def\eea{\end{eqnarray}}

\newcommand{\santi}[1]{{\color{black} #1}}

\begin{document}

\title{Magnetic fields: a tool for the study of organic solar cells}

\author{S. Oviedo-Casado${}^{1}$} 
\email{santiago.oviedo@upct.es}
\author{A. Urbina${}^{2}$, and J. Prior${}^{1,3}$}
\affiliation{${}^{1}$ Departamento de F{\'i}sica Aplicada, Universidad Polit{\'e}cnica de Cartagena, Cartagena 30202, Spain \\
${}^{2}$ Departamento de Electr{\'o}nica, Universidad Polit{\'e}cnica de Cartagena, Cartagena 30202, Spain \\
${}^{3}$Instituto Carlos I de F{\'i}sica Te{\'o}rica y Computacional, Universidad de Granada, Granada 18071, Spain}

\begin{abstract}
 Charge transfer in polymer devices represents a crucial, though highly inaccessible stage of photocurrent generation. 
In this article we propose studying the properties and behaviour of organic solar cells through the modification of photocurrent generation when an external magnetic field is applied. 
By allowing the parameters of our theoretical model not to be constrained to any specific material, we are able to show that not only a modest external magnetic field leads to a significant increase in photocurrent intensity, but also how such magnetic field can be used to study in detail the energy levels and transition rates within the polymer compound. Systematic exploration of key properties in organic composites thus can lead to highly optimised devices in which a magnetic field produces an enhancement in the efficiency of polymer solar cells.
\end{abstract}

\maketitle

\section{Introduction}

The ubiquitous goal in photovoltaic research is to improve the performance of existing materials and configurations, as nowadays the best polymeric photovoltaic (OPV) cells present an efficiency not 
surpassing 11\% \cite{EfficiencyTable,Dimitrov2014,Cao2012,Scharber2006}. Typical configurations for the active layer of these devices are conjugated polymers blended with fullerene derivatives 
\cite{Sheng2015,Park2009,Vandewal2009}, whose quantum efficiency, namely the ratio of absorbed photon to created exciton, raises in these compounds above 0.9 \cite{Bakulin2012,Gelinas2014a}. This 
means that there exists a necessity to bridge the gap existing between the initial high efficiency and the final photocurrent generated, and for that purpose the position of the energy levels --and 
in particular the charge transfer states-- has been proposed to be critical \cite{Rao2013a}. In addition, photocurrent generation is highly dependent on the recombination rates of the different 
energy levels, hence they are strongly responsible for the final efficiency of OPVs \cite{Deibel2010,Deotare2015}. Although the link between the relative position of intermediate energy level bands 
and global recombination rates is not yet well understood, it is clear that unravelling the inner dynamics of polymer compounds holds the key to fabricate improved polymer solar cells.


Intermediate stages in the process of photocurrent generation on polymer devices are mediated by electron-hole  (e-h) pairs which, contrary to what happens in other photovoltaic systems such as 
photosynthetic complexes, involve transport both of charge and energy \cite{Veldman2009}. Moreover, OPVs require the electron to be completely separated from the hole in order to generate current. In 
between the formation of an electron-hole pair and its eventual dissociation (or recombination), lies a series of ladder-like steps in which the electron progressively separates from the hole. These 
are denominated charge transfer (CT) states, and minimising recombination during the time the electron-hole pair is in this stage is crucial to maximise photocurrent generation 
\cite{Shockley1961,Vandewal2009,Deotare2015}. Studying charge transfer states is however difficult, since they are nearly dark and therefore not sensitive enough for spectroscopic techniques 
\cite{Vandewal2009}. On the other hand, charge transfer states involve spin 1/2 particles, susceptible to the influence of a magnetic field, which can be used to alter the properties of charge 
transfer states and thus modify the generated photocurrent --an easily measured outcome-- hence providing information about the different properties of the OPV, and  allowing to systematically 
search for optimal designs. In addition, applying a magnetic field could also lead to more efficient solar cells, leading polymer devices to be eventually commercially advantageous over solid state, 
semiconductor solar cells.


In this article, we complement and expand the results presented in \cite{Oviedo2017}, where although a magnetic field was shown to be able to influence the photocurrent generation process by 
enhancing triplet charge transfer state formation via an increased inter-system crossing, it was found that the required magnetic fields were too strong to be of commercial interest. Thus potential 
experimental applications were mentioned. In this article, not only we refine the model to show that a significant increase in photocurrent intensity can be obtained at lower magnetic fields and for 
realistic polymer materials, but we primarily focus on demonstrating with numerous examples the experimental possibilities offered by employing a magnetic field as a  tool to elucidate the 
different and optimal characteristics of real polymer compounds. This article is organised as follows: Section 2 introduces the OPV theoretical model that we 
use. Section 3 is devoted to present the results for an OPV in which the magnetic field produces appreciable changes in photocurrent intensity. In Section 4 the different spectroscopic possibilities and
applications of an external magnetic field are demonstrated in detail. We finish the article with some conclusions in Section 5.

\section{Model}
\label{model}

\begin{figure}
\centering
 \includegraphics[width=10cm]{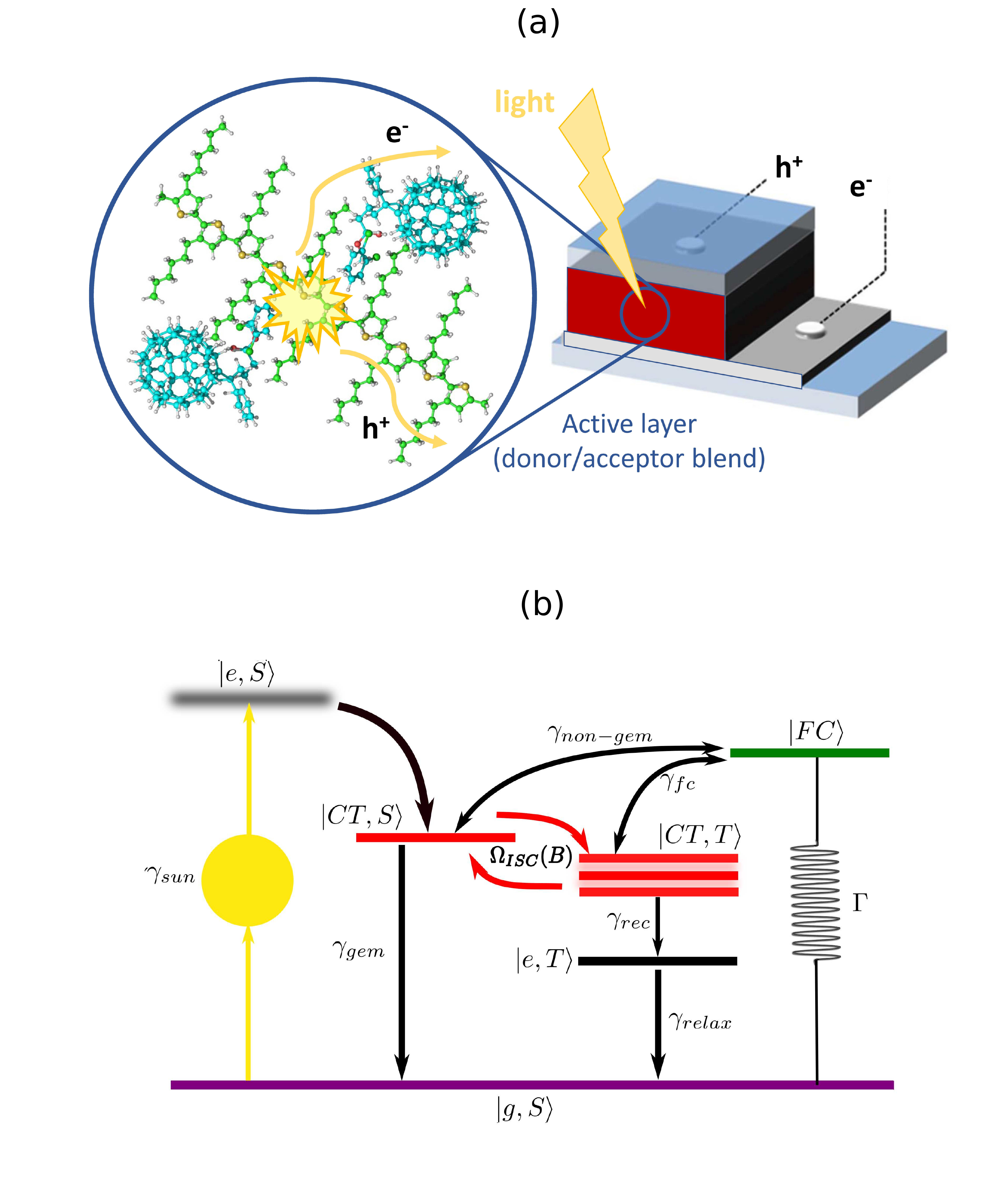}
 \caption{(a) shows an schematic organic solar cell structure, with a donor-acceptor layer composed of P3HT:PCBM, where incoming light induces electronic transitions to a higher state of a $\pi$ 
orbital in the donor. The Coulombically bound electron-hole pair rapidly migrates to the donor:acceptor interface where charge separation occurs and a net photocurrent is created, leading ultimately 
to electricity generation. On (b), our model solar cell and photocurrent generation process proposal is presented. Incoming photons create singlet exciton states that evolve into singlet CT states in 
hundreds of fs. Inter-system crossing converts coherently back and forth between singlet and triplet CTs, the later of which are split due to Zeeman effect caused by the interaction with an external 
magnetic field. CT states can either recombine to the ground state or dissociate to form free charges. Both processes are incoherent, \santi{represented by black, solid arrows whose relative 
width} represents the characteristic rates for the corresponding transitions. An impedance between the free charge and the ground state allows to 
study the I-V characteristics of the 
model.}
\label{Fig1}
\end{figure}

In order to fully understand the process of photogeneration in polymer solar cells, microscopic theoretical models are required; from photoexcitation, through charge separation, until the electron 
reaches the positive plate, as is schematically shown in the upper panel of Figure \ref{Fig1}, where we already glimpse the difficulty of coming up with an exact model, owing to 
the high number of degrees of freedom composing even the most simple polymer:fullerene blends, and the typically disordered structure that such materials present \cite{Noriega2013}. We propose here an 
heuristic approach to the problem, in which we identify the dynamically relevant independent degrees of freedom, and associate a quantum state to each of them, thus having an ansatz 
capable of qualitatively reproducing the behaviour of realistic OPVs which at the same time is able to provide information about the key processes occurring within the blend.

The model that we propose is depicted in the lower panel of Fig. \ref{Fig1}. An incoming photon generates an electron-hole pair (exciton $\ket{e,S}$) in the polymer, acting here as a donor. The 
electron rapidly migrates to the interface between donor and acceptor (fullerene in \ref{Fig1}(a)) where, being loosely bound to the original hole, it forms a charge transfer state ($\ket{CT,S}$). As 
this process occurs with near unity quantum efficiency, and in a much faster time-scale than all subsequent evolution (i.e $\approx$ 100 fs), we can regard it as instantaneous throughout 
the present study \cite{Gelinas2014a} and therefore, we will consider initial excitation to take place directly to a charge transfer state. Subsequent evolution sees the electron either dissociating 
completely, or recombining with the same hole (geminate) or a different one (non-geminate) into the ground state $\ket{g,S}$. If dissociation occurs photocurrent is generated  and we describe it as 
a free charge ($\ket{FC}$) state. It is the delicate balance between recombination and free charge generation what sets the global efficiency of the OPV. 

The initially generated e-h pairs conserve the spin of the ground state, i.e. they are in a singlet state and we have labelled them accordingly. However, the possibility of having triplet charge 
transfer states ($\ket{CT,T}$) is real, and leads to a different, complementary pathway of generating free charges. Moreover, as the ground state has spin zero, direct recombination of triplet CT 
states is forbidden \cite{Rao2013a} and has to happen through an auxiliary triplet exciton state ($\ket{e,T}$), hence the dissociation of free charges is in principle more efficient if it happens 
from triplet CT states.The necessary conversion from the initial singlet CT to the more convenient triplet CT --a process called inter-system crossing (ISC)-- occurs either 
via hyperfine and spin-orbit coupling or due to wavefunction overlap, all of them too small to be relevant in OPVs. The alternative is to enhance spin flips employing an external magnetic field 
\cite{Schulten1978,Frankevich1992,Cohen2009,Hontz2015}.

That an external magnetic field is able to enhance inter-system crossing has been controversial \cite{Deotare2015,Hontz2015}, owing mainly to the time-scales involved. For an external magnetic field 
to be able to induce spin flips the precession frequency induced by the magnetic field has to be faster than a) the time it takes for the spins to align in the direction of the magnetic field and b) 
the recombination rates for singlet CT states. Up to now, the time-scales studied were in the range of $\mu$s, where we are in the scenario where the magnetic field only aligns spins and potential ISC 
is either neutral (i.e. not observable) or negative, as it converts already existing triplet CTs into singlets. Crucially, our model follows those of Friend et al. and considers evolution of CT 
states in the nanosecond time-scale, where precession of electron spins and differences existing in the coupling of the electron and the hole to the external magnetic field (mainly due to a different 
g-factor) make the external magnetic field not only an important source but the main source of ISC. A combination of this effect with a slower triplet CT recombination pathway (w.r.t singlet CT) 
generates a situation in which the population of the triplet CT state is enhanced, thus leading to faster FC generation.

The Hamiltonian describing the model presented in Fig. \ref{Fig1}(b) is
\be
\begin{split}
\mathcal{H}_S &= E_0\ket{g,S}\bra{g,S} + E_s\ket{CT,S}\bra{CT,S} + E_t\ket{CT,T}\bra{CT,T} \\& + E_{et}\ket{e,T}\bra{e,T} + E_{FC}\ket{FC}\bra{FC} 
 + \mathcal{H}_m,
\end{split}
\label{Hamiltonian}
\ee
including all the relevant energy levels and where interaction with an external magnetic field is accounted for by
\be
\mathcal{H}_m = \sum_{T=0,\pm 1}\Bigl(2\mu_B g T B\ket{CT,T}\bra{CT,T} + \Omega_{ISC}(B)(\ket{CT,S}\bra{CT,T} + \ket{CT,T}\bra{CT,S})\Bigr),
\ee
\santi{with $\mu_B$ the Bohr magneton, g is the g-factor, T runs over the different triplet spin values and B is the magnetic field. The factor $\Omega_{ISC}(B)$ includes all possible sources of 
inter-system crossing, both internal (mainly hyperfine coupling) as well as the external magnetic field.} 

Incoherent transitions between energy levels, including recombination and charge dissociation are accounted for via Lindbladian terms of the form
\be
\mathcal{L}_{\alpha} = \gamma_\alpha \left[\sigma_\alpha \rho(t) \sigma_\alpha^\dagger - \frac{1}{2} \left\{ \sigma_\alpha^\dagger \sigma_\alpha,\rho(t) \right\}\right],
\label{Lindblad}
\ee
where each $\gamma_\alpha$ represents a different transition rate (see Fig. \ref{Fig1}(b)), and each operator $\sigma_\alpha$ describes the hopping between corresponding energy levels.

In this manuscript we are interested in obtaining a measurement of the performance of the corresponding solar cell. With this purpose, the intensity of the generated photocurrent at the steady state 
becomes the 
ideal quantity  to be analysed and eventually optimised. Such intensity is gauged as the amount of free charges that decay to the ground state via an artificial impedance simulated as an infinite 
particle reservoir with coupling rate 
$\Gamma$. With this definition, the photocurrent intensity is $I = e\rho_{FC,FC}\Gamma$ and the voltage difference is $V = E_{FC} - E_g + \frac{1}{\beta}ln\frac{\rho_{FC,FC}}{\rho_{gg}}$, with $e$ 
the electron 
charge, and \santi{$\beta$ = $\frac{1}{k_BT_{\Gamma}}$ with $k_B$ the Boltzmann constant and $T_{\Gamma}$ the corresponding impedance (here simulated as an environment) temperature}, which is room 
temperature throughout this manuscript. \santi{Here, the voltage arises from the energy released in equilibrating the system with an environment at temperature $T_\Gamma$, and is calculated through the 
derivative of the logarithm of the corresponding partition function}. The density matrix elements in these definitions are obtained from the steady state solution of the full 
Lindblad equation 
\be
\frac{d\rho}{dt} = -i\left[\mathcal{H}_S,\rho(t)\right] + \sum_\alpha \mathcal{L}_\alpha \left[\rho(t)\right],
\ee
whose initial state is the ground state and where solar-like excitation to the singlet CT is simulated as the interaction with a particle reservoir of high temperature photons with upwards transition rate of 
$\gamma_{sun} = 0.1$.

\section{Efficient solar cell in a magnetic field}
\label{section2}

In this section we present a state-of-the-art polymer solar cell subject to the influence of a magnetic field, demonstrating that with realistic parameters, a meaningful increase in both intensity 
and power outcome with relatively small magnetic fields can be achieved. 

\begin{figure}[!h]
\includegraphics[width=13cm]{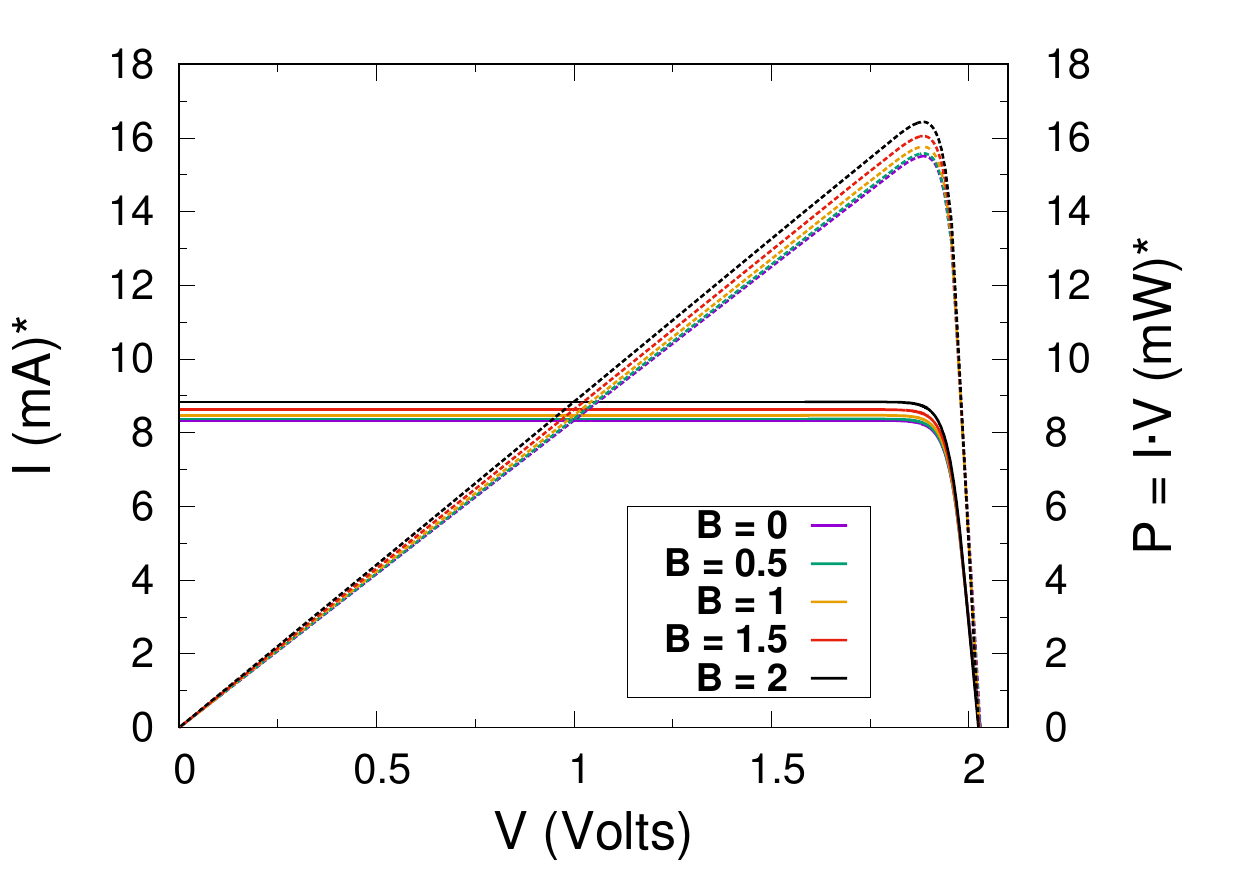}
\begin{minipage}{0.42\columnwidth}
\vspace{-14 cm}
\hspace{-4 cm}
\includegraphics[width=4.6cm]{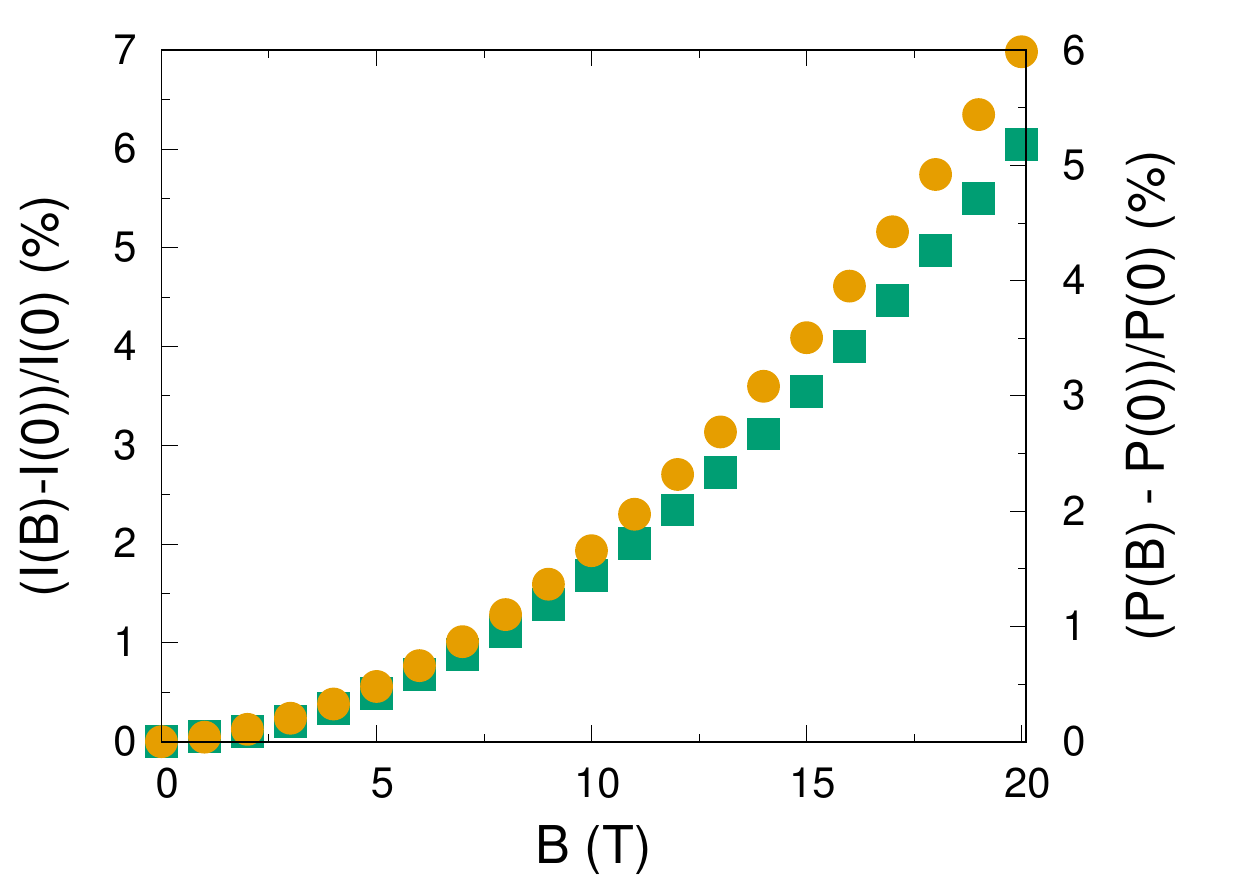}
\end{minipage}
\caption{Simulated I-V characteristic \santi{(continuous)} and power outcome \santi{(dashed)} of the model solar cell with singlet CT recombination rate of 1 ns while the triplet CT recombination 
rate is one order of magnitude slower. The different values of the external magnetic field (ranging here from zero to two Tesla) show how a moderate increase leads to enhancement in the performance of 
the OPV. The inset shows the percentage increase in intensity (dots) and power (squares) calculated at the point of maximum power. Geminate recombination rate \santi{$\gamma_{gem}$} is in this case 
0.1 ns, with non-geminate recombination rate \santi{$\gamma_{non-gem}$} being 10 ns, free charge generation rate \santi{$\gamma_{fc}$} is 1 ns and triplet exciton recombination rate 
\santi{$\gamma_{relax}$} is 10 ns. The asterisk in the intensity and power units means that we work with \santi{units normalised to a simulated solar-like illumination (see main text)}.}
\label{IVcurve}
\end{figure}

Figure \ref{IVcurve} shows the I-V characteristic for our solar cell model in the presence of a magnetic field ranging from zero to two Tesla, together with the generated power 
corresponding to such intensity, defined as P = I$\cdot V$. Both characteristics (I-V and P-V), which are representative of the OPV performance, present a significant increase in the presence of a 
magnetic field. In \cite{Oviedo2017} we employed 
the same model but with restringing parameters that matched those of a well known but scantily efficient polymer compound, such as the P3HT can be. Here instead, we allow for much more freedom in 
order to search for the ideal polymer material that maximises the enhancing effect of a magnetic field in photocurrent. Thus we choose a typical singlet CT recombination rate of one nanosecond 
\cite{Veldman2009}, while allowing for the triplet CT recombination rate to be one order of magnitude slower, hence favouring the protection of population in such triplet CT state, thus facilitating 
dissociation of charges into free charge states and photocurrent against recombination and losses. Note that even though we choose parameters in line with the desired effect we 
seek, these remain realistic, as it is now possible in frontier materials to achieve the effective freezing of the triplet CT states \cite{Rao2013a,FlatteImmense}, consequently turning off their 
recombination pathway. One can easily imagine how in such materials, employing a magnetic field should be --as we propose-- ideal to enhance the performance of polymer solar cells.

In addition, the inset in Figure \ref{IVcurve} shows the percentage increase in photocurrent intensity and power whose origin is the external magnetic field. Such percentage, calculated at the 
point of maximum power, represents an alternative, complementary measurement of the OPV performance. The magnitude of the intensity \santi{and power} shown in Fig. \ref{IVcurve} depends on the 
solar-like 
illumination \santi{$\gamma_{sun}$, which is an statistical measure of the number of incoming solar photons and provides} numerical values for the different parameters that are in such a way 
normalised to a standard ``size'' unit of the solar cell. \santi{We would like to emphasise that despite not having the more usual intensity per unit area typical from experimental results in solar 
cells,} the percentage increase is independent of this solar-like illumination, thus the 6\% increase at two Tesla represent a significant step forward in the quest of obtaining polymer solar 
cells that can 
outcompete commercially the classical semiconductor solar cells.

The quality of a solar cell can be measured by its filling factor (FF), which is defined through the maximum power attainable as $P_{mp} = I_{mp}\cdot V_{mp} = FF\cdot I_{SC}\cdot V_{OC}$, where 
$I_{SC}$ is the current at short circuit (i.e zero voltage) and $V_{OC}$ is the voltage at open circuit (i.e no current). We observe that for the model example presented in Fig. \ref{IVcurve} the 
filling factor is 0.9, and it increases slightly with the magnetic field, providing an independent measure of the beneficial effect that can be obtained from applying an external magnetic field. 
Such number is above the best FF found nowadays in real OPVs, which reach up to around 0.8 \cite{Guo2013}. The difference is easily explained in the simplicity of our model, which does not take 
directly into 
account factors such as electron mobility or structural disorder, which in real experiments always lower the quality of an OPV, and which are inevitable.

\section{Magnets as laboratory instruments for solar cell investigation}

After demonstrating that we can directly interfere in  the photocurrent generation process of an OPV, let us in this section introduce a different, complementary, and potentially very useful application of an external magnetic field. Namely its capacity as an experimental tool to access information about energy levels disposition and transport rates. Through the modification of the generated photocurrent, we will show in the subsequent paragraphs that, based on a model such as we presented in section \ref{model}  it is possible to infer certain properties of the OPV subject to the magnetic field. 

Information about transition rates and energy levels is usually obtained via spectroscopic techniques, such as pump-probe or 2D spectroscopy. These methods rely on the capacity of the energy states 
to absorb and emit radiation and thus, whenever the amount of states is too big or they lack optical strength being therefore dark, getting information via spectroscopy is challenging. On the other 
hand, employing a magnetic field to affect the photocurrent intensity provides an indirect measure that does not depend so heavily on the above mentioned properties of the states, thus being much 
more reliable for 
the case of OPVs. 

\subsection{Rates variations}

\begin{figure}
\includegraphics[width=12cm]{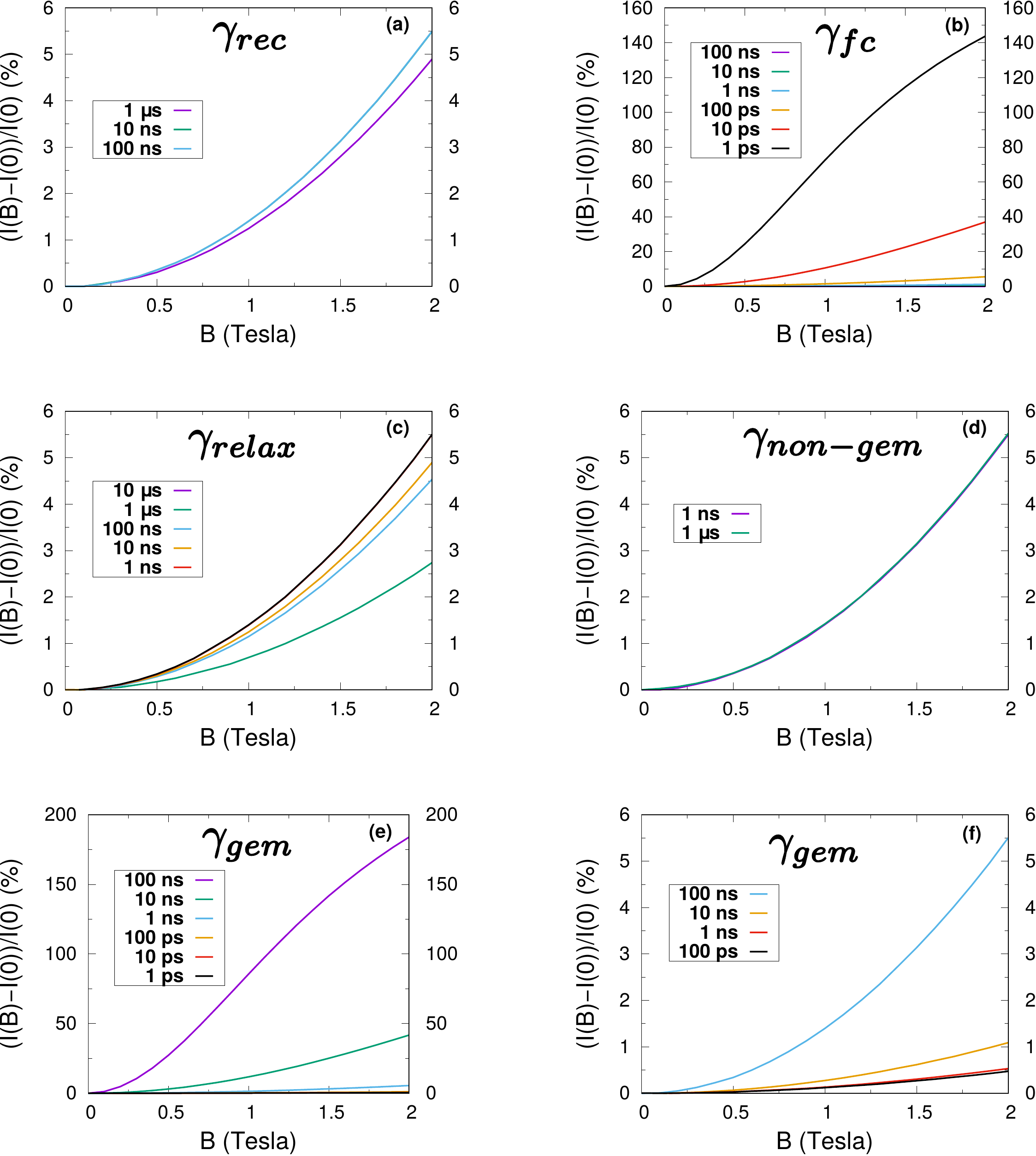}
\caption{From left to right and top to bottom, percentage variation with respect to 0 Tesla external magnetic field, on the intensity of the generated photocurrent for different incoherent 
transitions rates, labelled as in Fig. \ref{Fig1}(b). (a) shows variations of the triplet CT recombination rate. (b) is the variation of 
the FC generation rate. (c) presents the triplet recombination rate variation. (d) presents the variation of the non-geminate recombination rate, though no difference at all is appreciable, (e) displays the curves from different geminate recombination rates while (f) is intended to amplify the last four curves in (e). In all cases the rates which are kept constant while one of them is varied, acquire values in line with those presented in 
Fig. \ref{IVcurve}.}
\label{Rates}
\end{figure}

Firstly, we study the behaviour of the current intensity curves when the OPV is subject to a magnetic field between 0 and 2 Tesla, and test the differences appearing when the various transition rates in 
our model (see Fig. \ref{Fig1}(b)) change. Figure \ref{Rates} shows the resulting percentage variation on the intensity of photocurrent for different values of each transition rate. It is clear 
that for all except for the non-geminate recombination rate, it is possible to extract information from the variation of the intensity in the presence of a magnetic field. This latter case does not 
show an appreciable variation due to its relative independence from the ISC as it affects directly and in the same way both singlet and triplet CT, hence not depending on the population present in 
the triplet CT state, but rather only on the global population present in the FC state. 

It is clear from this analysis where the emphasis in design should be put. Both the free charge generation and geminate recombination hold the key to not only improved designs but also to obtain 
astounding increases in efficiency due to the presence of an external magnetic field (up to 200 \% in the case of the geminate recombination rate). Moreover, we find that the increment tends to 
saturate, thus stating that there is no troublesome infinite energy generation possibility, which is a double-check to the validity of our model. For the remaining rates, the variations range from 
moderate to non-existent in the case of the non-geminate recombination rate. Where no more values have been calculated is because the resulting curves where indistinguishable from the presented ones. 
The fact that each of the rates display different influence in the photocurrent intensity in the presence of an external magnetic field means that a multivariate analysis of experimental results from 
several different samples, should be able to discern and infer the different values of the incoherent rates for each of the samples. Thus it is demonstrated the power of an external magnetic field to 
easily characterise different materials and designs.

\subsection{Energy level variations}

\begin{figure}
\includegraphics[width=12cm]{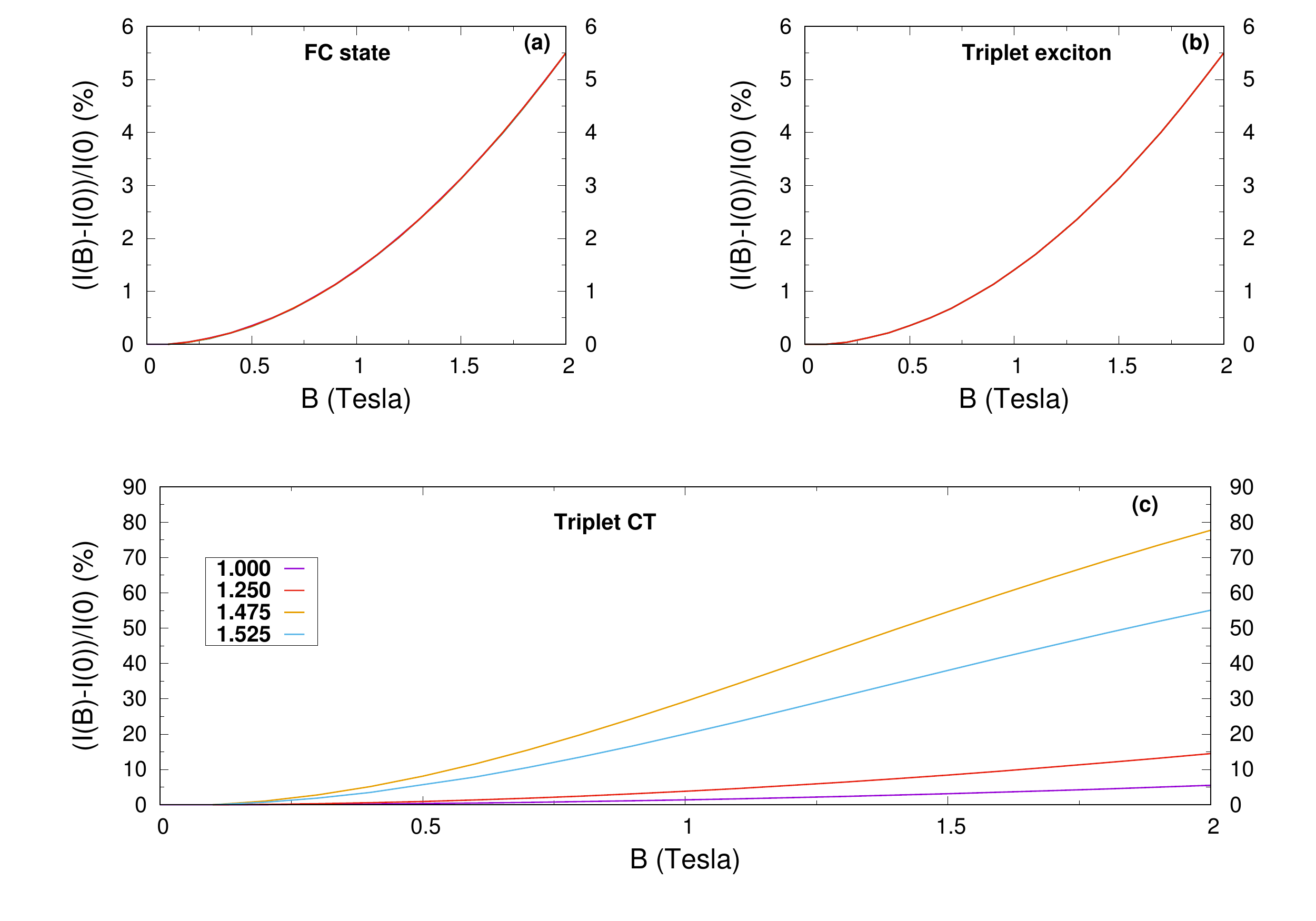}
\caption{Percentage variation on the intensity of the generated photocurrent, for different magnetic fields and changing the energies of different states. In (a) and (b), changing the energy of the 
free charge state and the triplet exciton state leads to no differences. In (c), changing the energy (in eV) of the triplet charge transfer state leads to a huge increase in the generated 
photocurrent 
whenever the triplet CT is close to resonance with the singlet CT (upper, yellow curve), while the effect is smaller if it is further from resonance or if it lies above the singlet CT.}
\label{Elevels}
\end{figure}

Only loosely bound e-h pairs are significantly affected by a magnetic field. This means that gaining information about exciton states using an external magnetic field is generally not viable. 
Moreover, Fig. \ref{Elevels}(top) shows that varying the energy of either the FC or the triplet exciton state has no influence on the intensity curve at any magnetic field. Thus in principle 
information of these states can only be accessed via modelling and through the variations in the corresponding transition rates, as we saw in the previous subsection. Furthermore it tells us that 
there is not much to gain from changing the relative position of these energy levels via clever engineering of OPVs. Unless the triplet exciton lays above the triplet CT state, in whose case 
decay of the triplet CT is strongly forbidden, which would lead to a great increase in intensity in the presence of a magnetic field, as can be concluded from the previous subsection. 

There is however much to be learned from the study of the variations that a magnetic field produces in intensity when the triplet CT state energy changes. We observe in Fig. \ref{Elevels} that a 
magnetic field is able to discern between different triplet CT energy configurations and, in addition, that the closer the triplet CT is to the singlet CT in energy --while being below-- the greater 
the enhancement in photocurrent intensity. However notice that complete resonance is not desired, as Zeeman splitting, while small enough, would render triplet states to lie above the singlet CT, 
thus making ISC transitions increasingly hindered.

\section{Conclusions}

Although historically an external magnetic field has been discredited as a potential useful tool in organic photovoltaics \cite{Deotare2015,Hontz2015}, we have demonstrated that its benefit is 
twofold. On the one hand, clever engineering of materials can lead to great increases in generated photocurrent and efficiency when the OPV is in the presence of a magnetic field. An external magnetic field is able to alter the balance existing between recombination and generation of free charges, inclining it more towards a greater generation, as can be seen in Fig. \ref{Rates}. And on the other hand, 
magnetic fields have the potential to become an essential part of the laboratory toolkit that analyses the properties and behaviour of different organic materials. Employing moderate magnetic fields 
can allow for access to information about each transition rate between the different energy levels, as well as the position of some of this energy levels, thus permitting an almost complete characterisation of the material subject to study. This means that the cost of designing new, improved materials from the existent ones can be greatly reduced, as now the information about the ideal properties the sample must have is readily available.


\section{Acknowledgement}
This research was undertaken with financial support from MINECO 
(SPAIN), including FEDER funds: FIS2015-69512-R  and No. ENE2016-79282-C5-5-R together with Fundaci{\'o}n S{\'e}neca (Murcia, Spain) Project No. 19882/GERM/15.

\section{Author contributions}

S.O.C. did the numerical simulations, J.P. and S.O.C. developed the theory. AU proposed the model and its application to PV technology. The manuscript was written by S.O.C. with
significant contributions from all the other authors.

\end{document}